\documentclass{IEEEtran}
\usepackage{bbm}
\usepackage{amssymb,amsmath,amsthm}
\usepackage{graphicx,epstopdf,psfrag,url,cite,xcolor,multirow,float}
\usepackage{caption-patch}
\usepackage[small,bf]{caption}
\usepackage[font=footnotesize]{subfig}

\usepackage{algorithm}
\usepackage{algorithmic}
\usepackage[super]{nth}
\usepackage{array}
\newcolumntype{C}[1]{>{\centering\let\newline\\\arraybackslash\hspace{0pt}}m{#1}}

\newtheoremstyle{mythm}% name
{\topsep}   % Space above1
{\topsep}   % Space below1
{\itshape}      % Body font
{0pt}       % Indent amount2
{\bfseries} % Theorem head font
{:}         % Punctuation after theorem head
{5pt plus 1pt minus 1pt}    % Space after theorem head3
{\thmname{#1}\thmnumber{ #2}\thmnote{ (#3)}}
\theoremstyle{mythm}

\newtheorem{proposition}{Proposition}

\newcommand{\bm}[1]{\boldsymbol{#1}}

\IEEEoverridecommandlockouts

\begin{document}

\title{Optimization-driven Deep Reinforcement Learning for Robust Beamforming in IRS-assisted Wireless Communications}

\author{Jiaye Lin\IEEEauthorrefmark{1}, Yuze Zou\IEEEauthorrefmark{2}, Xiaoru Dong\IEEEauthorrefmark{1}, Shimin Gong\IEEEauthorrefmark{1}, Dinh Thai Hoang\IEEEauthorrefmark{3} and Dusit Niyato\IEEEauthorrefmark{4}\\
\IEEEauthorblockA{
\IEEEauthorrefmark{1}School of Intelligent Systems Engineering, Sun Yat-sen University, China\\
\IEEEauthorrefmark{2}School of Electronic Information and Communications, Huazhong University of Science and Technology, China\\
\IEEEauthorrefmark{3}School of Electrical and Data Engineering, University of Technology Sydney, Australia\\
\IEEEauthorrefmark{4}School of Computer Science and Engineering, Nanyang Technological University, Singapore
}
}

%\author{
%Yongchang Deng\IEEEauthorrefmark{1}, Yuze Zou\IEEEauthorrefmark{2}, Shimin Gong\IEEEauthorrefmark{1}
%%, Wenqing Cheng\IEEEauthorrefmark{1}, Dinh Thai Hoang\IEEEauthorrefmark{4} and Dusit Niyato\IEEEauthorrefmark{5}
%\\
%\IEEEauthorblockA{
%\IEEEauthorrefmark{1}School of Intelligent Systems Engineering, Sun Yat-sen University, China\\
%\IEEEauthorrefmark{2}School of Electronic Information and Communications, Huazhong University of Science and Technology, China%\\
%%\IEEEauthorrefmark{4}School of Electrical and Data Engineering, University of Technology Sydney, Australia\\
%%\IEEEauthorrefmark{5}School of Computer Science and Engineering, Nanyang Technological University, Singapore
%}
%}

\maketitle
\thispagestyle{empty}

\begin{abstract}
Intelligent reflecting surface (IRS) is a promising technology to assist downlink information transmissions from a multi-antenna access point (AP) to a receiver. In this paper, we minimize the AP's transmit power by a joint optimization of the AP's active beamforming and the IRS's passive beamforming. Due to uncertain channel conditions, we formulate a robust power minimization problem subject to the receiver's signal-to-noise ratio (SNR) requirement and the IRS's power budget constraint. We propose a deep reinforcement learning (DRL) approach that can adapt the beamforming strategies from past experiences. To improve the learning performance, we derive a convex approximation as a lower bound on the robust problem, which is integrated into the DRL framework and thus promoting a novel optimization-driven deep deterministic policy gradient (DDPG) approach. In particular, when the DDPG algorithm generates a part of the action (e.g.,~passive beamforming), we can use the model-based convex approximation to optimize the other part (e.g.,~active beamforming) of the action more efficiently. Our simulation results demonstrate that the optimization-driven DDPG algorithm can improve both the learning rate and reward performance significantly compared to the conventional model-free DDPG algorithm.
\end{abstract}
%\begin{IEEEkeywords}
%Intelligent reflecting surface, Deep reinforcement learning, DDPG, robust optimization
%\end{IEEEkeywords}

\section{Introduction}

Recently, intelligent reflecting surface (IRS) has been introduced as a promising technology to improve energy- and spectrum-efficiency of wireless communications~\cite{19survey_renzo}. It is composed of a large array of passive scattering elements interconnected and individually controlled by an embedded IRS controller. The joint control of the complex reflecting coefficients for all scattering elements, i.e., passive beamforming, can enhance the signal strength to the receiver~\cite{joint_overview}. The IRS's passive beamforming along with the transceivers' transmission control are envisioned to further improve the network performance. An extensive survey in~\cite{gsm_survey} reveals that the IRS has already been applied in diverse network scenarios, serving different roles in wireless communications such as the ambient reflector, signal transmitter and even the receiver.

The use of IRS mainly aims at improving the transmission performance in terms of signal-to-noise ratio (SNR) or spectral efficiency~\cite{18pbf_rui1}, power consumption or energy efficiency~\cite{derrick}, and security enhancement~\cite{Yu2019Enabling}. The IRS can also be used to enhance wireless power transfer~\cite{lyubin}, mobile edge computing~\cite{computation}, and vehicular communications~\cite{veh-irs}. The performance maximization of IRS-assisted wireless systems is typically formulated as a joint optimization problem of the active and passive beamforming strategies, e.g.,~\cite{18pbf_rui1,derrick,Yu2019Enabling,lyubin,computation}. However, due to the non-convex problem structure, the solution methods are typically based on the alternating optimization (AO) framework with guaranteed convergence to sub-optimal solutions. Within each iteration of the AO framework, semidefinite relaxation (SDR) or convex approximation are usually required to optimize either the active or passive beamforming. As a heuristic approach, the performance loss of the AO method can not be known exactly and difficult to characterize precisely~\cite{derrick}. Besides, the optimization methods also suffer from a few practical difficulties. Firstly, the computational complexity of the AO method may increase significantly as the size of IRS's scattering elements becomes large. This makes it difficult for practical implementation in a dynamic radio environment. Secondly, a tractable formulation of beamforming optimization is in fact based on an inexact system modeling, which can be a simplification of the real system, for example, with perfect channel information, continuous and exact phase control. The problem reformulation or approximation further lead to a deviated solution far from the optimum. In general, we can expect that the model-based optimization methods only provide a lower performance bound of the original problem.

Different from the optimization methods, machine learning approaches are more robust against the uncertain system models and have also been applied to IRS-assisted wireless systems to realize the IRS's phase control. The authors in~\cite{19phase_chau} employ a deep neural network (DNN) to map the receiver's location to the IRS's optimal phase configuration in a complex indoor environment. Similarly, the authors in~\cite{unsupervised,dl-asu,dl-truly} use well-trained DNNs to make real-time predictions for the IRS's passive beamforming, which can achieve close-to-optimal performance with reduced time consumption or computational complexity compared to the SDR-based optimization methods. However, the offline training of DNNs relies on either an exhaustive search or the AO methods. The authors in~\cite{drl-miso} apply the deep deterministic policy gradient (DDPG) algorithm to maximize the received SNR of an IRS-assisted system by continuously interacting with the environment. The DRL approach is used in~\cite{drl-dusit} to enhance secrecy rate against multiple eavesdroppers. The authors in~\cite{drl-asu} implement the DRL agent at the IRS, which can observe the channel conditions and take actions based on the receiver's feedback. Though the DRL approach can learn the optimal strategy from scratch, it generally has a slow learning rate to converge.

In this paper, we propose a novel DRL approach with enhanced learning efficiency to minimize the transmit power of the access point (AP) in an IRS-assisted multiple-input single-output (MISO) system with uncertain channel conditions. We firstly formulate a robust power minimization problem by jointly optimizing the active and passive beamforming, and then we construct a Markov decision process (MDP) to solve it by learning from past experiences. To improve learning efficiency, we design the optimization-driven DDPG algorithm that integrates the model-based optimization into the framework of a model-free DDPG algorithm. In particular, when the DDPG algorithm generates a part of the action, a model-based optimization module can be used to find the other part of the action very efficiently. By solving an approximate convex problem, the optimization module also provides a achievable lower bound on the original robust problem, which guides the DDPG algorithm to search for an optimal action more efficiently. Our simulation results reveal that the optimization-driven DDPG algorithm not only speeds up the learning rate but also reduces the AP's transmit power significantly compared to the conventional DDPG algorithm.

\section{System model}\label{sec:model}

\begin{figure}[t]
	\centering
	\includegraphics[width=0.5\textwidth]{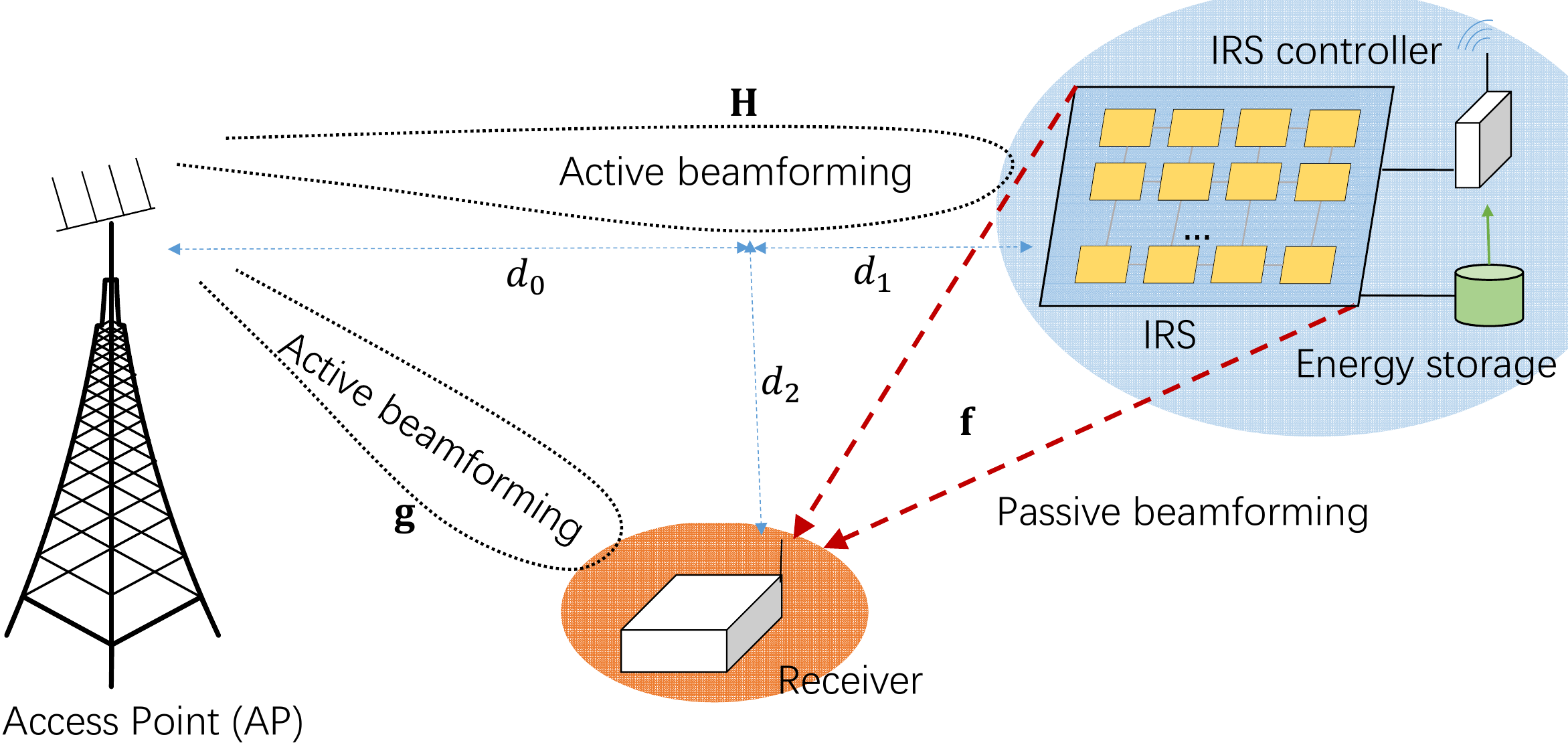}
	\caption{IRS-assisted MISO system.}\label{fig_system_model}
\end{figure}

As shown in Fig.~\ref{fig_system_model}, the IRS with $N$ reflecting elements assists the information transmissions from the $M$-antenna AP to the single-antenna receiver. The AP-receiver, AP-IRS and IRS-receiver complex channels are denoted by ${\bf g}\in \mathbb{C}^{M\times 1}$, ${\bf H}\in \mathbb{C}^{M\times N}$ and ${\bf f}\in \mathbb{C}^{N\times 1}$, respectively. We assume that each reflecting element can set a phase shift $\theta_n\in[0,2\pi]$ and its magnitude $\rho_n\in[0,1]$ to reflect the incident RF signals.

\subsection{SNR and Energy Budget}
Let $\bm{\Theta}=\text{diag}(\rho_1 e^{j\theta_1},\ldots,\rho_N e^{j\theta_N})$ denote the IRS's passive beamforming, where $\text{diag}(\cdot)$ denoting the diagonal matrix given the diagonal vector. Hence the IRS-assisted equivalent channel from the AP to the receiver is given by $\hat{\bf g}={\bf g}+{\bf H}{\bf \Theta}{\bf f}$, where ${\bf H}= [{\bf h}_1,\ldots,{\bf h}_N]$ denotes the channel matrix from the AP to the IRS. Let ${\bf w} \in\mathbb{C}^{M \times 1}$ denote the AP's beamforming vector and $s$ be the complex symbol with unit transmit power. The received signal at the receiver is thus $y=\hat{\bf g}^H{\bf w}s+\nu_d$, where the superscript denotes the conjugate transpose and $\nu_d$ denotes the Gaussian noise with zero mean and normalized variance. Therefore, the received SNR can be characterized as
\begin{equation}\label{eq_SNR}
\gamma({\bf w}, {\boldsymbol{\Theta}})=\lVert ({\bf g}+{\bf H}{\bf \Theta}{\bf f})^H{\bf w}\rVert^2,
\end{equation}
which depends on the AP's active beamforming ${\bf w}$ and the IRS's passive beamforming ${\bf \Theta}$.

%\subsection{IRS's Power Consumption Model}
%Given the AP's transmit beamforming, the incident signal at the IRS is ${\bf x}={\bf H}^H{\bf w}s$.

We assume that each tunable chip of the reflecting element is also equipped with an energy harvester that is able to harvest RF energy from the AP's beamforming signals. By tuning the magnitudes of reflecting coefficients $\bm{\rho}\triangleq [\rho_1, \ldots, \rho_N ]^T$, a part of the incident signal power is reflected to the receiver, while the other part is fed to the energy harvester. Hence, the parameter $\bm{\rho}$ is also called as the power-splitting (PS) ratio. To maintain the IRS's operations, the total harvested energy has to meet the IRS's total power consumption. That is, $\eta\sum_{n} (1 - \rho_n^2) \lVert{\bf h}^H_n{\bf w}\rVert^2 \geq N\mu$, where $\eta$ represents the power harvesting efficiency and ${\bf h}_n$ denotes the channel from the AP to the $n$-th reflecting element. The IRS's power consumption is given by $N\mu$, where $\mu$ is the power consumption of a single scattering element~\cite{Huang2018Large}.

\subsection{Channel Uncertainty Model}

We assume that the direct channel ${\bf g}$ from the AP to the receiver can be estimated accurately by the active receiver. In particular, the HAP can send a known pilot information to the receiver with fixed transmit power. Meanwhile, the IRS switches off its reflecting elements. The channel ${\bf g}$ can be recovered at the receiver based on the received signal samples.

However, by using the passive scattering elements, the channels ${\bf H}$ and ${\bf f}$ have to be estimated at either the HAP or the receiver by overhearing the channel response. We assume that the channel ${\bf H}$ is subject to estimation errors, i.e.,~${\bf H} = \bar{\bf H} + {\bf \Delta}_{\bf h}$, where $\bar{\bf H}$ denotes the averaged estimate and ${\bf \Delta}_{\bf h}$ denotes the error estimate of the channel from the HAP to the IRS. The error estimate ${\bf \Delta}_{\bf h}$ has limited power density, and thus we can define the uncertainty set $\mathbb{U}_{{\bf h}}$ for the AP-IRS channel ${\bf H}$ as follows:
\begin{equation}\label{equ_uncertain_h}
{\bf H} \in \mathbb{U}_{\bf h} \triangleq \{{\bf H}= \bar {\bf H} + {\bf \Delta}_{\bf h}: \textbf{Tr}({\bf \Delta}_{\bf h}^H{\bf \Delta}_{\bf h}) \leq \delta_{\bf h}^2 \},
\end{equation}
where $\textbf{Tr}(\cdot)$ represents the trace operation and $\delta_{\bf h}$ denotes the power limit of error estimate ${\bf \Delta}_{\bf h}$. The estimation of the IRS-receiver channel ${\bf f}$ has to be bundled with the AP-IRS channel ${\bf H}$ and performed at the receiver by overhearing the mixture of signals from the HAP and the IRS's reflections. By rewriting the cascaded AP-IRS-receiver channel as ${\bf H}_{\bf f} \triangleq\text{diag}({\bf f}) {\bf H}= [f_1 {\bf h}_1, f_2 {\bf h}_2,\ldots, f_N{\bf h}_N]$, similar to~\eqref{equ_uncertain_h} we can define the uncertainty for ${\bf H}_{\bf f}$ as follows:
\begin{equation}\label{equ_uncertain_f}
{\bf H}_{\bf f} \in \mathbb{U}_{\bf f} \triangleq \{{\bf H}_{\bf f}= \bar {\bf H}_{\bf f} + {\bf \Delta}_{\bf f}: \textbf{Tr}({\bf \Delta}_{\bf f}^H{\bf \Delta}_{\bf f}) \leq \delta_{\bf f}^2 \},
\end{equation}
where $\delta_{\bf f}$ denotes the power limit of the error estimate ${\bf \Delta}_{\bf f}$ for the reflecting channel ${\bf H}_{\bf f}$. The average channel estimate $\bar {\bf H}_{\bf f}$ and the power limit $\delta_{\bf f}$ are assumed to be known in advance by channel measurements.

\section{Robust Active and Passive Beamforming}

%\subsection{Robust Counterpart and Reformulations}

We aim to minimize the AP's transmit power, denoted as $||{\bf w}||^2$, by jointly optimizing the active and passive beamforming strategies, subject to the IRS's power budget constraint and the receiver's SNR requirement. Considering a practical case that all the reflecting elements have the same PS ratio ${\rho}$, we can decompose the optimization of $\rho$ and the phase vector $\boldsymbol{\theta}=[e^{j\theta_1},\ldots,e^{j\theta_N}]^T$. As such, we can rewrite the IRS-assisted channel by $\hat{\bf g} =  {\bf g}+{  \rho} {\bf H}_{\bf f} {\bm\theta}$ and then simplify robust power minimization problem as follows:
\begin{subequations}\label{prob_robust}
\begin{align}
\min_{{\bf w}, \boldsymbol{\theta},\rho}~&~\lVert {\bf w} \rVert^2\label{robust_obj}\\
s.t.~&~\lvert ({\bf g}+{  \rho} {\bf H}_{\bf f} {\bm\theta})^H{\bf w} \rvert^2 \geq \gamma_1,\quad \forall \, {\bf H}_{\bf f}\in\mathbb{U}_{\bf f},\label{SNR}\\
~&~\eta (1 - \rho^2) \lVert{\bf H}^H{\bf w}\rVert^2 \geq N\mu,\quad \forall \, {\bf H}\in\mathbb{U}_{\bf h},\label{energy}\\
~&~\rho\in(0,1) \text{ and } \theta_n \in (0, 2\pi) \quad \forall \, n\in\mathcal{N}.\label{con_power_simp}
\end{align}
\end{subequations}
%The constraints in~\eqref{SNR} and~\eqref{energy} define the receiver's worst-case SNR requirement and the IRS's worst-case power budget constraint, respectively.
The first difficulty of the non-convex problem~\eqref{prob_robust} lies in that the PS ratio $\rho$ is coupled with the phase vector ${\bm \theta}$. Another difficulty comes from the semi-infinite constraints in~\eqref{SNR}-\eqref{energy}, which have to hold for any channel error estimate in the uncertainty set. In the sequel, we consider solving problem~\eqref{prob_robust} by using a learning-based and model-free approach that can tolerate inaccuracies in modeling and online decision-making.
%In the sequel, we first present a simple heuristic to decompose the optimization of $\rho$ and ${\bm \theta}$. Then, we reformulate the semi-infinite constraints in~\eqref{SNR}-\eqref{energy} and present a convex approximation to problem~\eqref{prob_robust}.

\subsection{Deep Reinforcement Learning Approach}

Deep reinforcement learning (DRL) is a combination of deep neural networks (DNNs) and reinforcement learning (RL). It aims at solving MDP problems with large action and state spaces that are difficult to handle by conventional RL approaches~\cite{sutton2018reinforcement}. The MDP framework can be defined by a tuple $\{\mathcal{S},\mathcal{A},\mathcal{P},\mathcal{R}\}$. $\mathcal{S}$ represents the system state space denoting the set of observations of the network environment. $\mathcal{A}$ denotes the set of actions. The state transition probability $\mathcal{P}$ denotes the distribution of the next state ${\bf s}_{t+1}\in \mathcal{S}$ given the action ${\bf a}_t\in\mathcal{A}$ taken in the current state ${\bf s}_t$. It is typically uncertain to the agent and has to be learnt during the decision making process. The immediate reward $\mathcal{R}:\mathcal{S}\times \mathcal{A}\rightarrow \mathbb{R}$ provides the quality evaluation $r_t({\bf s}_t, {\bf a}_t)$ of the state-action pair $({\bf s}_t, {\bf a}_t)$. It also drives the search for the best policy to maximize the long-term reward $V({\bf s}) =  \mathbb{E}\left[ \sum_{t=0}^{\infty} \gamma^{t} r_t({\bf s}_t, {\bf a}_t) | {\bf s} \right]$, where $\gamma\in[0,1]$ denotes the discount factor.

\subsubsection{MDP Reformulation}

The most straightforward DRL solution to problem~\eqref{prob_robust} is to design a DRL agent at the AP, which jointly decides the AP's transmit beamforming and the IRS's passive beamforming strategies, based on the observed state ${\bf s}_t\in\mathcal{S}$ and the knowledge learnt from past experience $\mathcal{H}_t\triangleq \{ \ldots, {\bf s}_{t-1}, {\bf a}_{t-1}, r_{t-1}, {\bf s}_{t} \}$. The system state ${\bf s}_t=({\bf c}_t, {\bf o}_t)$ includes the channel information and the indicator of outage events in the past signal transmission periods. The channel information is denoted by the set ${\bf c}_t\triangleq\{{\bf g}_t, {\bf H}_t, {\bf H}_{{\bf f},t}\}_{t\in\mathcal{T}}$, where $\mathcal{T}\triangleq\{t-T+1,\ldots,t-1,t\}$ denotes $T$ consecutive past transmission periods. We can easily extract the averaged channel estimates $\bar {\bf H}$ and $\bar{\bf H}_{\bf f}$ from set ${\bf c}_t$ by averaging over $T$ consecutive channel estimations. The power limit $\delta_{\bf h}$ (or $\delta_{\bf f}$) of the error estimate corresponding to the channel ${\bf H}$ (or ${\bf H}_{\bf f}$) can be also obtained similarly. In the system state, we also record the outage events ${\bf o}_t \triangleq [o_1,o_2,\ldots,o_T]^T$ in the past $T$ transmission periods. For each $t\in\mathcal{T}$, let $o_t = 1$ denote an outage event if~\eqref{SNR} or~\eqref{energy} does not hold, which implies the infeasibility of problem~\eqref{prob_robust}. Given the current state ${\bf s}_t$, the action ${\bf a}_t \triangleq({\bf w}_t, {\bm \theta}_t, \rho_t)$ includes the AP's active beamforming ${\bf w}_t$ and the IRS's passive beamforming, characterized by $(\rho_t, {\bm \theta}_t)$, which are both continuous decision variables. The AP's transmit power will be wasted once outage events happen. Hence, we can define the immediate reward as the energy efficiency, i.e.,~the successfully transmitted data over the AP's power consumption:
\begin{equation}\label{equ-reward}
r_t({\bf s}_t,{\bf a}_t) = \mathbb{E}[(1-o_t)|({\bf g}_t+{  \rho} {\bf H}_{{\bf f},t} {\bm\theta}_t)^H{\bf w}_t|^2||{\bf w}_t||^{-2}].
\end{equation}
The expectation is taken over the past $T$ transmission periods. It is clear that the reward is inversely proportional to the AP's transmit power when the inequalities hold in~\eqref{SNR}-\eqref{energy}, and otherwise the reward becomes zero when outage happens.

\subsubsection{DDPG for Continuous Control}
RL provides a solution to find the optimal policy $\pi^*: \mathcal{S} \to \mathcal{A}$ that maps each state ${\bf s}_t\in\mathcal{S}$ to an action ${\bf a}_t \in \mathcal{A}$ such that the value function $V({\bf s})$ is maximized. With small and finite state and action spaces, the optimal policy can be obtained by the Q-learning algorithm~\cite{sutton2018reinforcement}, i.e., the optimal action ${\bf a}_t^* = \arg\max_{{\bf a}\in\mathcal{A}} Q({\bf s}_t, {\bf a})$ on each state is to maximize the Q-value function, and then we update the Q-value by the difference between the current Q-value and its target value $y_t$:
\[
Q_{t+1}({\bf s}_t,{\bf a}_t) = Q_t({\bf s}_t,{\bf a}_t) + \tau_t \Big[ y_t - Q_t({\bf s}_t, {\bf a}_t)\Big],
\]
where $\tau_t$ is the step-size and the target $y_t$ is evaluated by
\begin{equation}\label{equ_target}
y_t =  r_t({\bf s}_t, {\bf a}_t) + \gamma \max_{{\bf a}_{t+1}} Q_t({\bf s}_{t+1}, {\bf a}_{t+1}).
\end{equation}
%It is clear from the above procedures that the update of Q-value is the main task in training.
For a small size of discrete state and action spaces, the Q-value for each state and action can be stored in a table and updated in each decision epoch. However, the Q-learning algorithm becomes unstable when the state and action spaces are very large~\cite{drl-hoang}. Instead, the deep Q Network (DQN) algorithm uses DNN with the weight parameter $\boldsymbol{\omega}_t$ as the approximator for the Q-value function. The input to the DNN is the current state ${\bf s}_t$ and output is the expected action ${\bf a}_t$. The DNN parameter $\boldsymbol{\omega}_t$ is updated in each decision epoch to output a better approximation of the Q-value. This can be achieved by training the DNN to minimize the loss function:
\begin{equation}\label{equ_loss}
\ell(\boldsymbol{\omega}_t)=\mathbb{E}\left[(y_{i} -Q_{t}({\bf s}_i, {\bf a}_i| \boldsymbol{\omega}_t))^{2}\right].
\end{equation}
One of the main advantages of using DRL lies in that we can learn from past experiences efficiently by using a group of historical transition samples $({\bf s}_i, {\bf a}_i, r_i, {\bf s}_{i+1})\in\mathcal{M}_t$, namely, a mini-batch $\mathcal{M}_t$, to train the DNN at each decision epoch. The expectation in~\eqref{equ_loss} is taken over all samples in the mini-batch $\mathcal{M}_t$ and the target $y_i$ is evaluated by~\eqref{equ_target} for each sample.

The DQN algorithm can be extended to solve optimization problems in a continuous action space. Besides a DNN approximator for the Q-value function, the DDPG algorithm also uses a DNN with the parameter ${\bm v}$ to approximate the policy function~\cite{lillicrap2015continuous}. The DNN training aims at updating the parametric policy $\pi_{\bm v}$ directly in a gradient direction to improve the estimation of value function, which is given as follows:
\begin{equation}\label{equ_ddpg_value}
J({\bm v}) = \sum_{{\bf s} \in \mathcal{S}} d({\bf s}) \sum_{{\bf a} \in \mathcal{A}} \pi_{\bm v}({\bf a} \vert {\bf s}) Q({\bf s},{\bf a}|\boldsymbol{\omega}),
\end{equation}
where $d({\bf s})$ denotes the stationary state distribution corresponding to the policy $\pi_{\bm v}$ and $Q({\bf s},{\bf a}|\boldsymbol{\omega})$ is the Q-value approximated by the DNN with parameter $\boldsymbol{\omega}$. By deterministic policy gradient theorem~\cite{sutton2018reinforcement}, the gradient of~\eqref{equ_ddpg_value} is given by
\begin{equation}\label{equ_gradient}
\nabla_{\bm v} J({\bm v}) = \mathbb{E}_{{\bf s} \sim d({\bf s})} [ \nabla_{\bf a} Q({\bf s},{\bf a}|\boldsymbol{\omega}) \nabla_{\bm v} \pi_{\bm v}({\bf s}) \rvert_{{\bf a}=\pi_{\bm v}({\bf s})}],
\end{equation}
which can be performed efficiently by sampling the historical trajectories. The policy gradient in~\eqref{equ_gradient} motivates the actor-critic framework, which updates two sets of DNN parameters $({\bm v},\boldsymbol{\omega})$ separately. The actor network updates the policy parameter ${\bm v}$ in gradient direction as follows:
\[
{\bm v}_{t+1} = {\bm v}_t + \alpha_{v} {\nabla_{\bf a} Q({\bf s}_t, {\bf a}_t|\boldsymbol{\omega}_t) \nabla_{\bm v} \pi_{\bm v}({\bf s}) \rvert_{{\bf a}_t=\pi_{\bm v}({\bf s})}}.
\]
The critic network updates the Q-network as follows:
\[
\boldsymbol{\omega}_{t+1} = \boldsymbol{\omega}_t + \alpha_{\omega} \delta_t \nabla_{\boldsymbol{\omega}} Q({\bf s}_t, {\bf a}_t|\boldsymbol{\omega}_t),
\]
where $\delta_t  = y_t - Q({\bf s}_t, {\bf a}_t|\boldsymbol{\omega}_t)$ denotes the temporal-difference (TD) error between $Q({\bf s}_t, {\bf a}_t|\boldsymbol{\omega}_t)$ and its target value $y_t$. Two constants $\alpha_{v}$ and $\alpha_{\omega}$ are viewed as step-sizes. The training of the critic network is also performed by sampling a mini-batch from the experience replay memory. To ensure better convergence, the target $y_i$ for critic network is given by
\begin{equation}\label{equ_target2}
y_i = r_i({\bf s}_i, {\bf a}_i) + \gamma Q({\bf s}_{i+1},\pi({\bf s}_{i+1}|{\boldsymbol{v}'_t})|\boldsymbol{\omega}'_t).
\end{equation}
Here the DNN parameters $({\boldsymbol{v}'_t}, \boldsymbol{\omega}'_t)$ of the target networks are delayed copy of $({\boldsymbol{v}_t}, \boldsymbol{\omega}_t)$ from the online networks. Given the current state ${\bf s}_i$, the action is determined by the policy $\pi_{{\bm v}_t'}$ and then the Q-value is estimated by the DNN with parameter ${\bm \omega}_t'$.
%\begin{subequations}\label{softUpdate}
%\begin{align}
%\boldsymbol{\omega}'_{t+1} & = \tau \boldsymbol{\omega}_t + (1-\tau)\boldsymbol{\omega}'_t, \\
%\boldsymbol{\vartheta}_{t+1}'  &= \tau \boldsymbol{\vartheta}_t + (1-\tau)\boldsymbol{\vartheta}_t'.
%\end{align}
%\end{subequations}

%\section{Optimization-driven DRL Framework}
%\section{Optimization-driven Hierarchical Deep Reinforcement Learning Approach}

%\subsection{Hierarchical DDPG (H-DDPG) Framework}

\subsection{Optimization-driven Learning Strategy}

The conventional DDPG approach estimates the target value $y_t$ by the immediate reward $r_t({\bf s}_t, {\bf a}_t)$ and the target Q-network with the parameter $\boldsymbol{\omega}'_t$, as shown in \eqref{equ_target2}. To ensure better convergence performance, the target Q-network is evolving from the online Q-network by the following rule:
\begin{equation}\label{equ_update}
\boldsymbol{\omega}'_{t+1} = \tau \boldsymbol{\omega}_t + (1-\tau)\boldsymbol{\omega}'_t,
\end{equation}
where $\boldsymbol{\omega}_t$ denotes the DNN parameter of the online Q-network and $\tau$ is a step size. This implies strong coupling between the two Q-networks and may lead to slow learning rate. The main drawbacks can be understood from the following aspects.
\begin{itemize}
  \item Random initialization: Both the online and target Q-networks can be initialized randomly. In the early stage of learning, the DNN approximations of Q-value can be far from its optimum and thus probably misleading the learning process. Hence, the DDPG algorithm practically requires a long warm-up period to train both Q-networks.
  \item Inaccurate reward estimation: The evaluation of the immediate reward $r_t({\bf s}_t, {\bf a}_t)$ is based on the output of the critic network with non-optimal parameter, especially in the early stage of learning. The inaccurate reward estimation can be also far from its real value.
  \item Sensitive parameter setting: The choice of parameter $\tau$ to update $\boldsymbol{\omega}'_{t+1}$ in~\eqref{equ_update} is problematic. A small value of $\tau$ can stabilize but also slow down the learning, while a large value of $\tau$ implies strong correlation between the online and target Q-networks, which may result in the fluctuations of learning performance or even divergence.
\end{itemize}

In this following part, we aim to stabilize and speed up the learning process by estimating the target value $y_t$ in a better-informed and independent way.

\subsubsection{Merge Model-free and Model-based Estimations}

%\subsubsection{Optimization-driven Learning Framework}
Based on the DDPG framework, in the $t$-th decision epoch the actor network outputs the action ${\bf a}_t = (\rho_t, {\bf w}_t, {\bm \theta}_t)$ and the target Q-network produces an estimation of the target $y_t$, which can be very far from its optimum in the early stage of learning. This motivates us to use a model-based optimization method to estimate a lower bound on the target value $y_t$ based on partial system information. Specifically, we can divide the action ${\bf a}_t$ into two parts, i.e.,~the scalar $\rho_t$ and two vectors $({\bf w}, {\bm \theta})$. Given the PS ratio $\rho_t$, we can solve the robust problem~\eqref{prob_robust} by optimizing the active and passive beamforming $({\bf w}, {\bm \theta})$. This significantly simplifies the solution to problem~\eqref{prob_robust}, and hence we can easily find a lower bound on the target value $y_t$. Let $y_t'$ denote the lower bound on the target value and $({\bf w}', {\bm \theta}')$ be the optimized beamforming solution. We envision that the model-based optimization can provide a better-informed target, i.e.,~$y_t'>y_t$, especially in the early stage of learning.

\begin{figure}[t]
\centering
\includegraphics[width=0.5\textwidth]{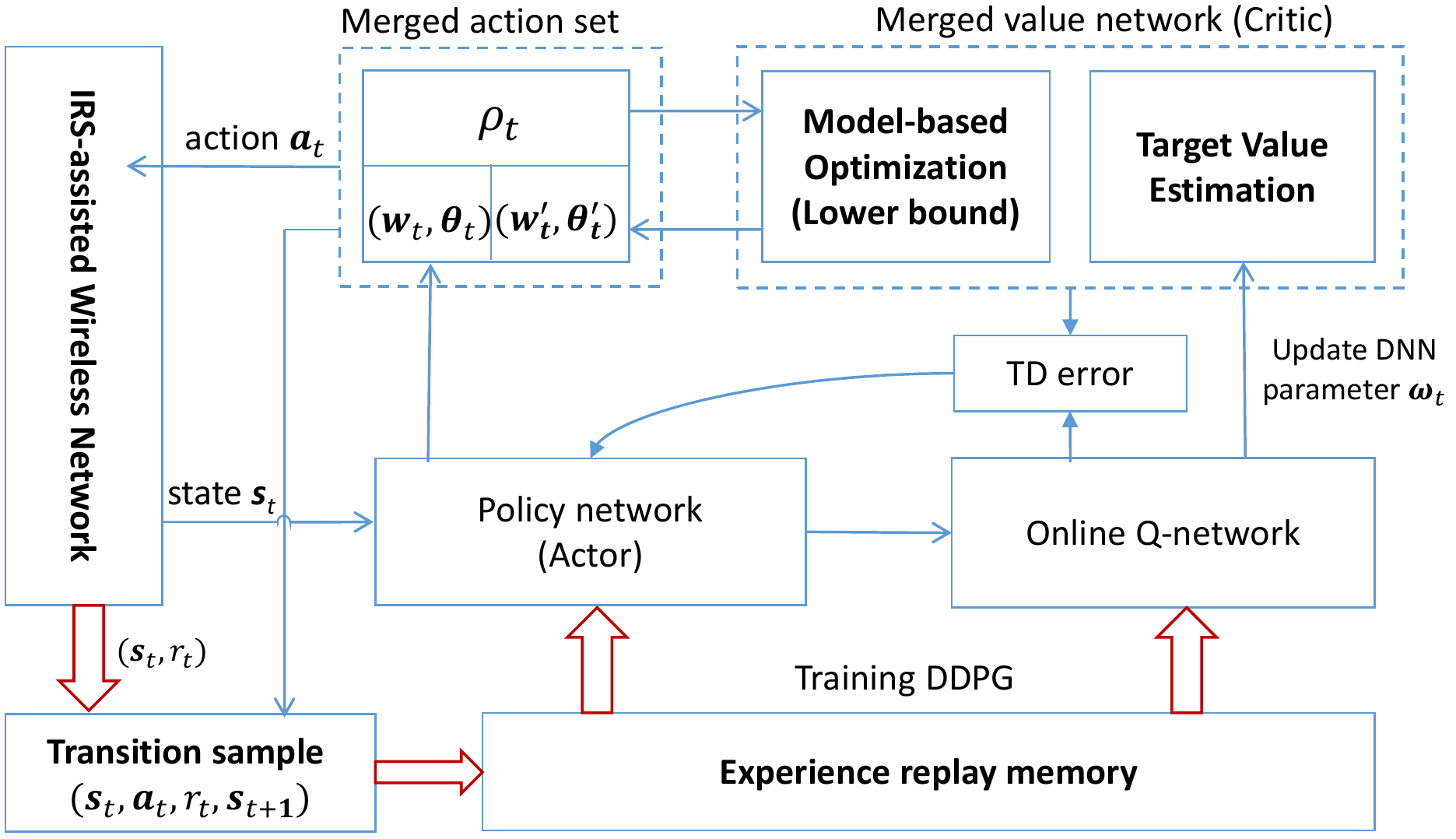}
\caption{The optimization-driven DDPG framework for active and passive beamforming in an IRS-assisted MISO downlink system.}\label{fig_hddpg}
\end{figure}

The flow chart of the proposed optimization-driven DDPG algorithm is shown in Fig.~\ref{fig_hddpg}. The actor and critic networks firstly generate the action and value estimates independently, respectively. Then, we fix $\rho_t$ and feed it into the model-based optimization module, which evaluates a lower bound $y_t'$ by solving an approximate problem of~\eqref{prob_robust}. Two target values $y_t$ and $y_t'$ can be merged by simply adopting the larger one in the following learning process. That is, for $y_t' > y_t$, we use $y_t'$ as the target value for the following DNN training and simultaneously update the beamforming strategy $({\bf w}'_t, {\bm \theta}'_t)$ in the action, i.e.,~${\bf a}_t = (\rho_t,{\bf w}'_t, {\bm \theta}'_t)$. We may also have $y_t' < y_t$ when the learning becomes more stable. In this case, we follow exactly the output of the actor network. The integration of the model-based optimization and the model-free learning may help the DDPG algorithm to adapt faster in the early stage. Moreover, the optimization-driven target value $y_t'$ is independent of the critic network. This implies that the target value $y_t'$ can be more stable than the output $y_t$ of the target Q-network during the training of online Q-network. Such a decoupling between the online Q-network and its target can reduce the performance fluctuation in training and thus it is expected to stabilize the learning in a shorter time.

\subsubsection{Model-based Approximation}
Till this point, we aim at finding an efficiently tractable approximation of problem~\eqref{prob_robust} to derive the lower bound on $y_t$. With a fixed $\rho$, we first determine a simple heuristic solution to the phase vector ${\bm \theta}$ and then focus on the optimization of the AP's transmit beamforming by reformulating a convex approximation of problem~\eqref{prob_robust}. The intuition behind our solution lies is that the IRS-assisted channel ${\rho} {\bf H}_{\bf f} {\bm\theta}$ can be aligned with the direct channel ${\bf g}$ for a large-size IRS. That is, we can always find a phase vector ${\bm \theta}$ such that ${\bf H}_{\bf f} {\bm \theta}=\kappa {\bf g}$, where $\kappa \in \mathbb{R^+}$ is a scalar constant. This allows us to simplify the constraint in~\eqref{SNR}.

Given the phase vector ${\bm \theta}$ and the channel gain $\kappa$, we then focus on the optimization of the AP's active beamforming in problem~\eqref{prob_robust}, which is tightly coupled with the uncertain channel matrices ${\bf H}_{\bf f}$ and ${\bf H}$. Fortunately, we have the following equivalence to the worst-case constraints in~\eqref{SNR}-\eqref{energy}.
\begin{proposition}\label{prop_snr}
Given the IRS's passive beamforming $(\rho, {\bm \theta})$, the constraints in~\eqref{SNR} and~\eqref{energy} have the following equivalent reformations, respectively:
\begin{subequations}\label{equ_cvx}
\begin{align}
&\left[\begin{matrix}
	\rho^2\left({\bm \theta}{\bm \theta}^H \otimes{\bf W}\right) + t {\bf I}_{MN} & \alpha\rho({\bm \theta} \otimes {\bf W}){\bf g}\\
	\alpha\rho {\bf g}^H ({\bm \theta}\otimes {\bf W})^H &\alpha^2{\bf g}^H{\bf W}{\bf g}-\gamma_1 - t \delta_{\bf f}^2
\end{matrix}\right] \succeq 0,\label{equ_cvx_snr}\\
&
\left[\begin{array}{cc}
	{  {\bf W}_c+ {\tau}{\bf I}_{MN},} &  {\bf W}_c \text{vec}(\bar{\bf H})\\
	{  \text{vec}(\bar{\bf H})^H{\bf W}_c,} & {  \bar \gamma_0 -\frac{N\mu}{\eta(1-\rho^2)}-\tau {\delta_{\bf h}^2}}
	\end{array}\right]	{  \succeq 0},\label{equ_cvx_energy}
\end{align}
\end{subequations}
for some $t\geq 0$ and $\tau \geq 0$, where $\alpha\triangleq(1+\rho \kappa_{m})$, ${\bf W}_c\triangleq {\bf I}_N \otimes{\bf W}$, and $ \bar \gamma_0 \triangleq  \text{vec}(\bar{\bf H})^H{\bf W}_c \text{vec}(\bar{\bf H})$. ${\bf I}_{MN}$ is the identity matrix with size $MN$. The semidefinite matrix ${\bf W}$ is a rank-one relaxation of ${\bf w}{\bf w}^H$, i.e.,~${\bf W} \succeq {\bf w}{\bf w}^H$.
\end{proposition}
The detailed proof for Proposition~\ref{prop_snr} is relegated to our online technical report~\cite{deng2020robust}. It is clear that with a fixed $\rho$ Proposition~\ref{prop_snr} transforms the semi-infinite constraints in~\eqref{SNR} and~\eqref{energy} into linear matrix inequalities in terms of ${\bf W}$ and the non-negative auxiliary variables $(t, \tau)$. Till now, we can easily derive a achievable lower bound on problem~\eqref{prob_robust} by the following convex problem:
\begin{equation}\label{prob_robust_1}
\min_{{\bf W}\succeq {\bf 0}, t\geq 0, \tau\geq 0}\{ {\bf Tr}({\bf W}) :\, \eqref{equ_cvx_snr} \text{ and } \eqref{equ_cvx_energy}\},
\end{equation}
which can be solved efficiently by the interior-point algorithms. The linear beamforming~${\bf w}$ can be retrieved by eigenvalue decomposition if the matrix solution ${\bf W}$ to~\eqref{prob_robust_1} is of rank one. Otherwise we can extract an approximate rank-one solution via Gaussian randomization~\cite{SDR}. Once we determine ${\bf w}$, we can update the energy efficiency as $({\bf g} +{\rho}_t \bar{\bf H}_{{\bf f},t} {\bm\theta}_t)^H{\bf w}_t|^2||{\bf w}_t||^{-2}$, which serves as the model-based evaluation of the target $y'_t$.

\section{Numerical Results}
In the simulation, we consider fixed network topology to verify the learning performance of the optimization-driven DDPG algorithm. As shown in Fig.~\ref{fig_system_model}, the distances in meters are given by $d_0 = 1$, $d_1= 1$, and $d_2= 2$. The signal propagation satisfies a log-distance model with the path loss at the reference point given by $L_0 = 30$~dB and the path loss exponent equal to 2. The energy harvesting efficiency is set as $\eta = 0.8$. In the following, we firstly demonstrate the learning performance of the proposed algorithm and then study the impact of different parameters on the AP's transmit power.

\begin{figure}[t]
  \centering
  \subfloat[AP's transmit power]{\includegraphics[width=0.25\textwidth]{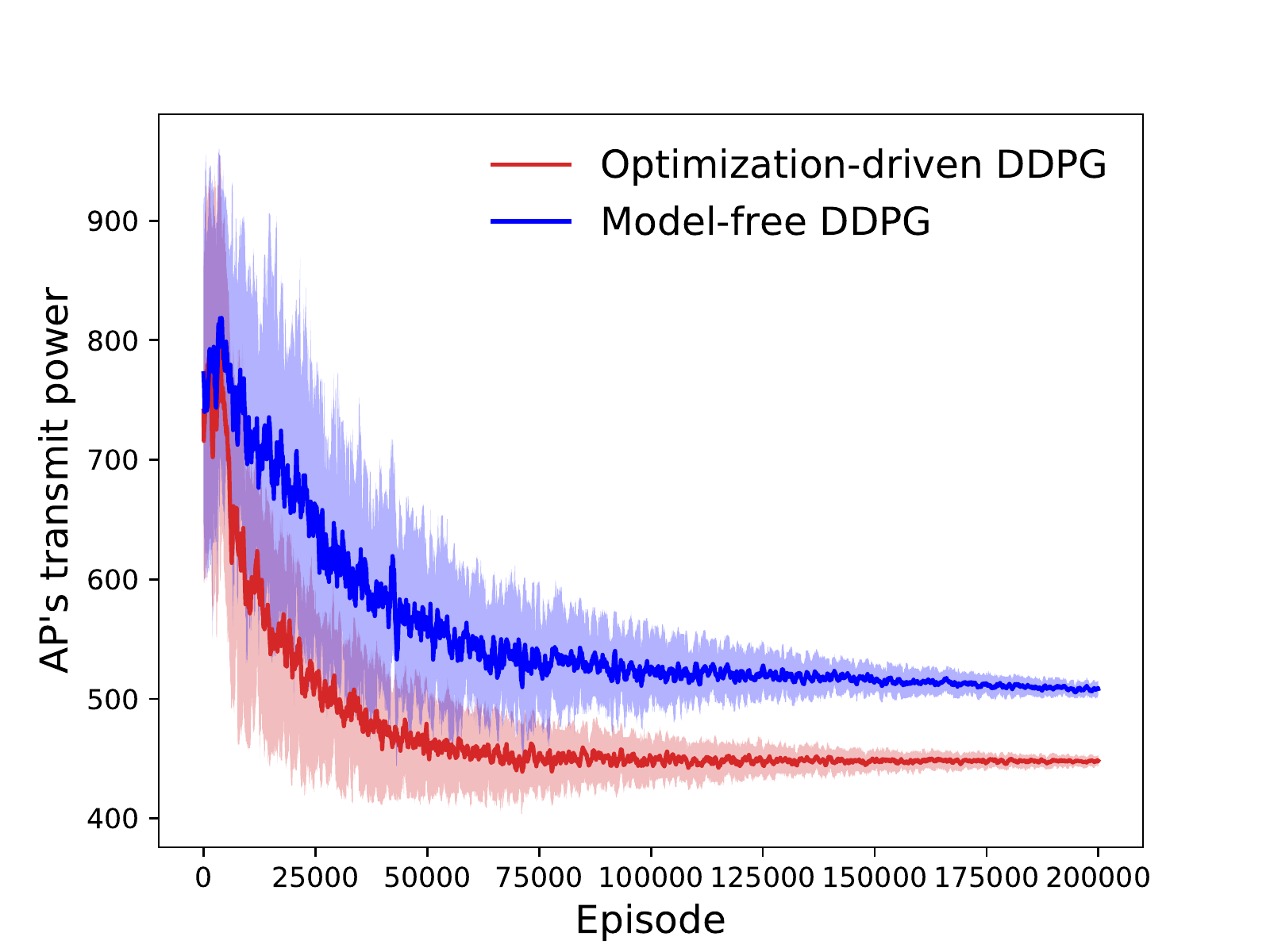}}
  \subfloat[Magnitude of reflection ]{\includegraphics[width=0.25\textwidth]{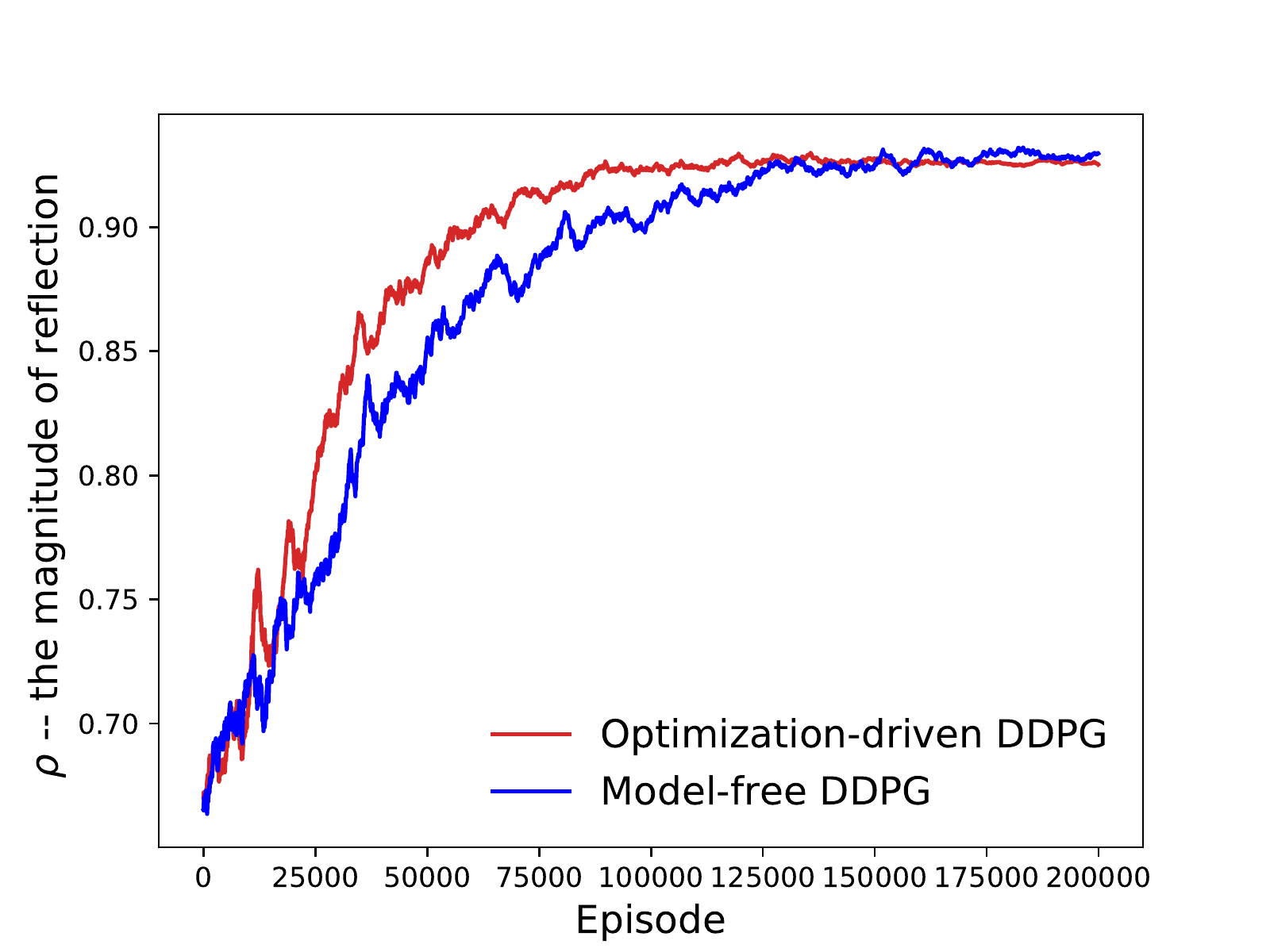}}
  \caption{Dynamics of the AP's transmit power and the magnitude of IRS's reflection coefficients in two DDPG algorithms. The solid line denotes the median of 50 repetitions and the shaded regions in different colors cover 10th to 90th percentiles.}\label{fig_conv}
  %\vspace{-0.5cm}
\end{figure}

Figure~\ref{fig_conv}(a) demonstrates the dynamics of the AP's minimum transmit power during the learning of the proposed \emph{Optimization-driven DDPG} algorithm. We also compare it to the conventional DDPG algorithm, denote as \emph{model-free DDPG} in Fig.~\ref{fig_conv}. The common observation is that the AP's transmit powers in both cases decrease gradually during the training process and converge eventually at two stable values. However, the optimization-driven DDPG converges faster than the model-free DDPG algorithm and also achieves a significant performance improvement in terms of the AP's transmit power. The reason is that the optimization-driven DDPG uses a better-informed estimation for the target value to guide its search for the optimal policy, which can achieve faster learning rate in the early stage of training compared to the model-free DDPG that learns from scratch. Furthermore, in Fig.~\ref{fig_conv}(b) we record the magnitude $\rho$ (the PS ratio) of the IRS's reflection coefficient in two algorithms. It is clear that the PS ratio $\rho$ increases gradually and converges to the same value within 200k episodes. The comparison in Fig.~\ref{fig_conv}(b) also reveals that the optimization-driven DDPG algorithm achieves a faster convergence rate compared to that of the model-free DDPG algorithm. Besides, we can observe from Fig.~\ref{fig_conv} that the learning curves are more stable in the proposed optimization-driven DDPG algorithm. The reason is that the model-based target estimation $y_t'$ is independent or decomposed from the online Q-network.

%This is consistent with the result shown in Fig.~\ref{fig1}, which verifies the advantages of the \emph{Optimization-driven DDPG} algorithm. %Figure 2 show the difference between the two algorithm. $\rho$ in Model-free DDPG converges the final value at a slower and more volatile rate, while Optimization-driven DDPG converges variables significantly faster than Model-free DDPG. Because the result of the optimization can be given a reference value for the DDPG part of the action, to help the three action variables that DDPG needs to optimize converge at a faster speed, thereby obtaining a better reward at a faster speed.

%\begin{figure}
%	\centering
%	\includegraphics[width=0.45\textwidth]{fig1-power}
%	\caption{Dynamics of the AP's transmit power in two DDPG algorithms. The solid line denotes the median of 50 repetitions and the shaded regions in different colors cover 10th to 90th percentiles.}\label{fig1}
%\end{figure}
%
%
%
%\begin{figure}
%	\centering
%	\includegraphics[width=0.45\textwidth]{fig2-rho}
%	\caption{Dynamics of the PS ratio $\rho$ in two DDPG algorithms.}\label{fig2}
%\end{figure}

\begin{figure}[t]
  \centering
  \subfloat[Impact on the SNR requirement]{\includegraphics[width=0.25\textwidth]{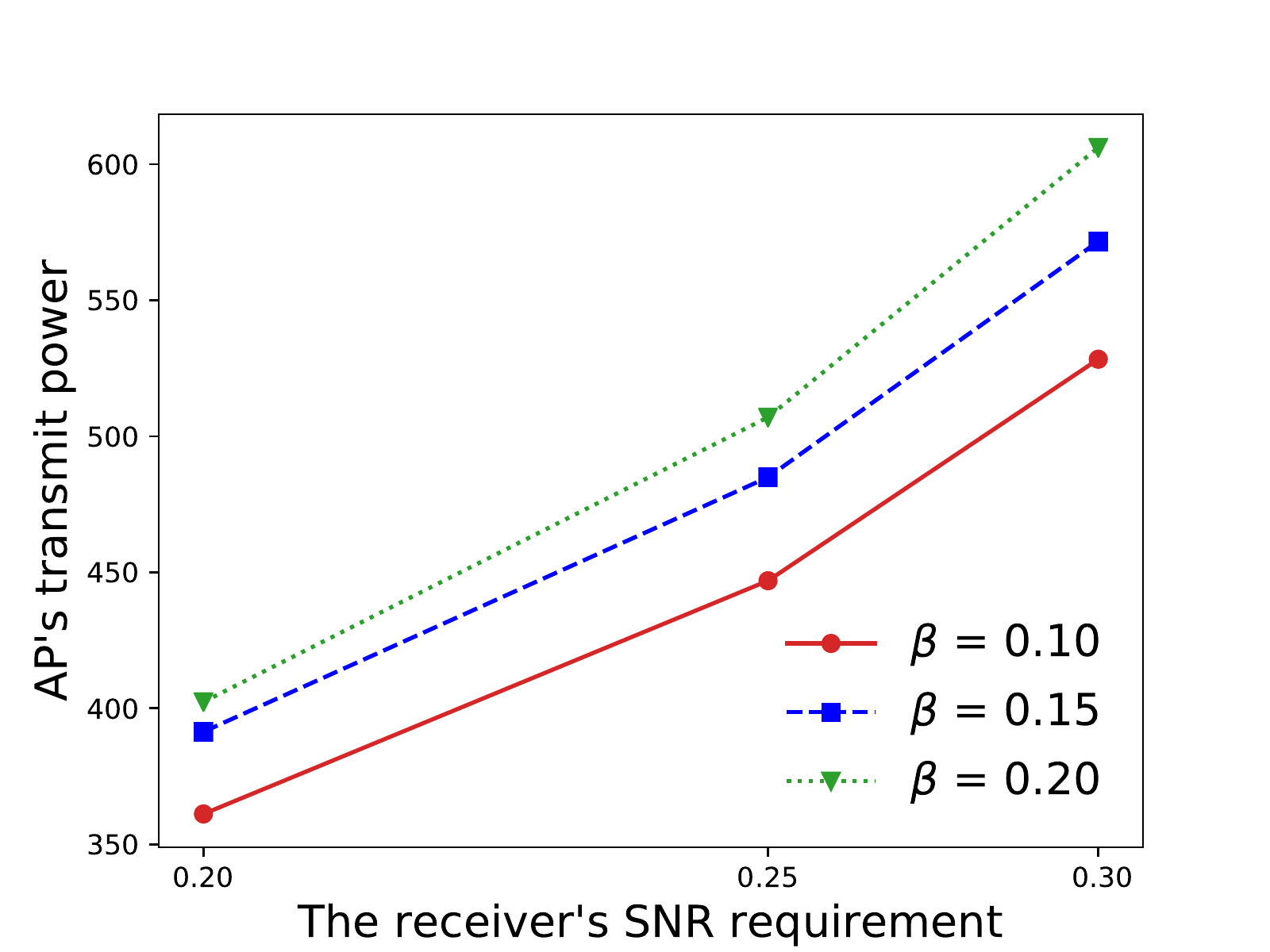}}
  \subfloat[Impact on the IRS's size]{\includegraphics[width=0.25\textwidth]{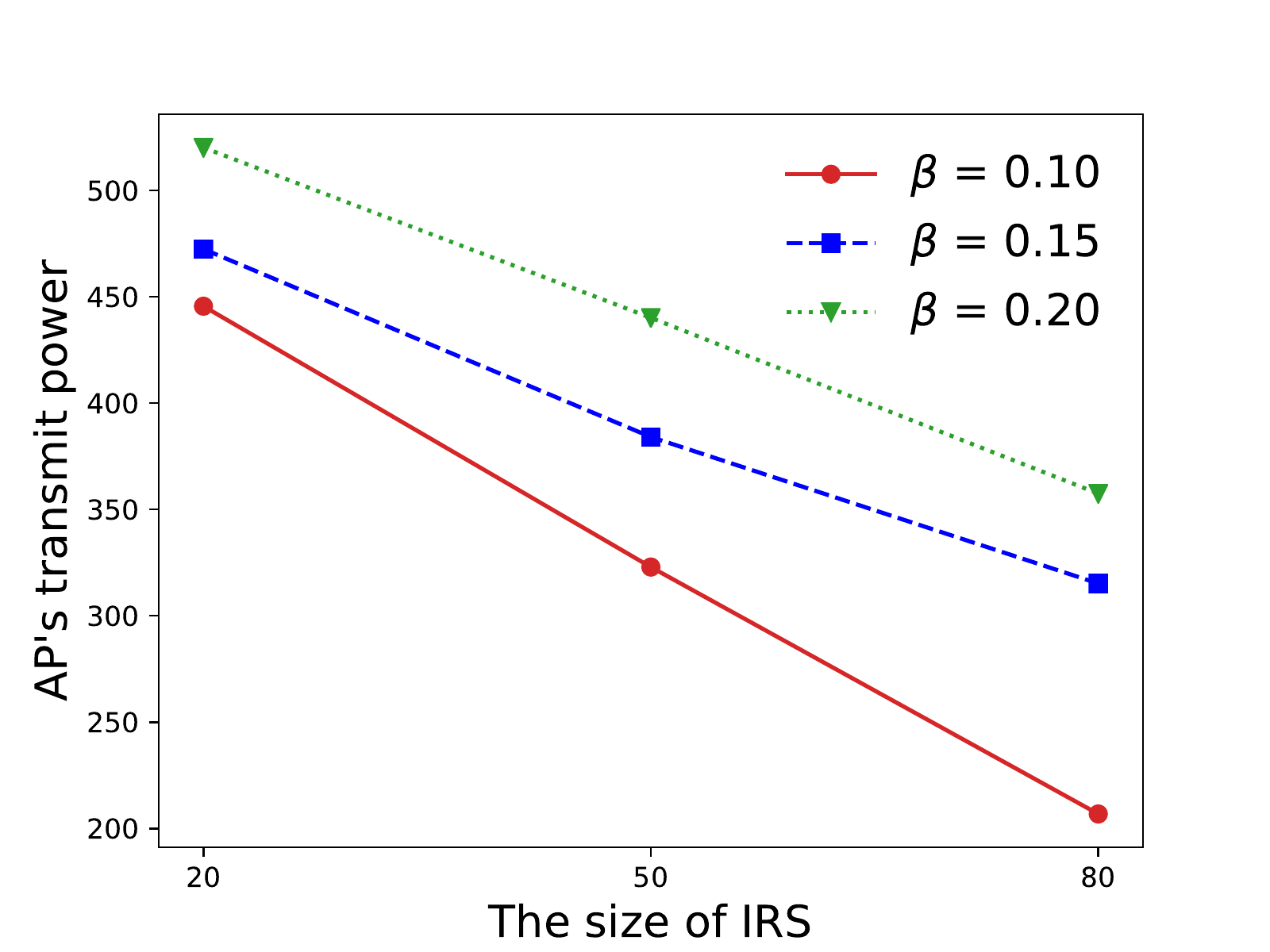}}
  \caption{AP's transmit power increases with the receiver's SNR requirement and decreases with the size of IRS's scattering elements.}\label{fig_param}
  %\vspace{-0.5cm}
\end{figure}

In the following, we evaluate the performance of the proposed optimization-driven DDPG against different parameters, including a) the receiver's worst-case SNR requirement $\gamma_1$ in problem~\eqref{prob_robust}, b) the size $N$ of the IRS's scattering elements, and c) the channel's uncertainty level $\beta$, defined by the ratio between error estimate and the mean channel estimate, i.e.,~$\beta_{\bf h} \triangleq \delta_{\bf h}^2 / \textbf{Tr}(\bar{\bf H}\bar{\bf H}^H)$ and $\beta_{\bf f} \triangleq \delta_{\bf f}^2 / \textbf{Tr}(\bar{\bf H}_{\bf f}\bar{\bf H}_{\bf f}^H)$. In the simulation, we set $\beta_{\bf h} = \beta_{\bf f} = \beta$. Fig.~\ref{fig_param}(a) shows that the AP's minimum transmit power increases with the increase in the receiver's SNR requirement $\gamma_1$. For the same SNR requirement, the AP's transmit power also increases with a higher uncertainty level $\beta$. Such an increase in the AP's transmit power can be viewed as the price of robustness to ensure worst-case performance guarantee. In Fig.~\ref{fig_param}(b), we show the dynamics of the AP's transmit power with a different number of the IRS's scattering elements. Though a higher uncertainty level leads to increased transmit power, such a negative effect can be alleviated by using a larger-size IRS.

To show the scalability of the DRL approach, we compare the average running time of an SDR-based optimization method to solve problem~\eqref{prob_robust} and that of the optimization-driven DDPG algorithm. The results are shown in Table~\ref{tab-compare}. Note that the SDR-based optimization method has a polynomially increasing computational complexity in terms of the problem size $M\times N$. Hence, it becomes very time-consuming with a large number of the IRS's scattering elements. On the contrary, the optimization-driven DDPG algorithm achieves nearly $\mathcal{O}(1)$ complexity in the size $M\times N$, as it only relies on a light-weight convex optimization to derive a lower bound for more efficient learning. Such a low complexity makes it very suitable for practical deployment, especially with a large-size IRS and a large number of the AP's transmit antennas.

\begin{table}[t]
	\centering
	\caption{Comparison of running time between the SDR-based optimization and the optimization-driven DDPG algorithms}\label{tab-compare}
	\begin{tabular}{|c|C{3cm}|C{3cm}|}
		\hline
		\multirow{2}{*}{$M\times N$} & \multicolumn{2}{c|}{Running time (in milliseconds)} \\ \cline{2-3}
							  & SDR &  DRL \\ \hline
		40 & 6.22 & 12.03 \\
		60 & 12.91 & 12.41 \\
		80 & 19.27 & 12.82 \\
		100 & 28.67 & 13.25 \\
		200 & 136.20 & 15.19 \\ \hline
	\end{tabular}
\end{table}

\section{Conclusions}
In this paper, we propose a novel DRL approach to solve a robust power minimization problem in an IRS-assisted MISO system, subject to the receiver's worst-case data rate requirement and the IRS's worst-case power budget constraint. Different from the conventional DDPG algorithm, we devise the optimization-driven DDPG algorithm that combines the benefits of both model-free learning and model-based optimization. Our simulation results demonstrate that the new DDPG algorithm can guide the search for the joint active and beamforming more efficiently. Both the learning rate and reward performance can be improved significantly compared to the conventional model-free DDPG algorithm.

\bibliographystyle{IEEEtran}

\bibliography{irs-drl}

\end{document}